# PubPeer and Self-Correction of Science: Male-Led Publications More Prone to Retraction

Abdelghani Maddi[1], Emmanuel Monneau[2], Catherine Guaspare-Cartron[3], Floriana Gargiulo[4], Michel Dubois[5]

[1] abdelghani.maddi@cnrs.fr
ORCID : 0000-0001-9268-8022
GEMASS – CNRS – Sorbonne Université, 59/61 rue Pouchet 75017 Paris, France.
[2] manu_monneau@hotmail.com
GEMASS – CNRS – Sorbonne Université, 59/61 rue Pouchet 75017 Paris, France.
[3] catherine.guaspare@cnrs.fr
GEMASS – CNRS – Sorbonne Université, 59/61 rue Pouchet 75017 Paris, France.
[4] floriana.gargiulo@cnrs.fr
GEMASS – CNRS – Sorbonne Université, 59/61 rue Pouchet 75017 Paris, France.
[5] michel.dubois@cnrs.fr
ORCID : 0000-0001-6872-9525
GEMASS – CNRS – Sorbonne Université, 59/61 rue Pouchet 75017 Paris, France.

## Abstract

This article has a dual objective. Firstly, it aims to investigate whether gender diversity in publications reviewed on Pubpeer has an impact on the (non)retraction of those publications. Secondly, it seeks to analyze the reasons for retractions and examine if there are disparities in retractions based on male-female collaborations. To achieve this, the study utilized a sample of 93,563 publications discussed on Pubpeer spanning the period from 2012 to 2021. The findings reveal that among the reviewed publications, 5% (4,513) were retracted. The concentration index and regression results indicate that publications authored solely by men or led by male authors are 20% to 29% more likely to be retracted compared to those authored solely by women.

Regarding the reasons for retractions, the results show that regardless of gender, authors, when working alone, are more prone to engaging in activities such as fake peer review or plagiarism. Women are more concentrated in image manipulation and data errors, while men are more involved in article duplication. Furthermore, the results demonstrate an inverse relationship between the number of authors and retractions, suggesting that a higher number of authors may facilitate better publication control and reduce the temptation for misconduct.

## Acknowledgments
Data available from The Center for Scientific Integrity, the parent nonprofit organization of Retraction Watch, subject to a standard data use agreement. The authors would like to thank the Pubpeer Foundation for authorizing the collection and use of data associated with the operation of their platform.

## Competing interests
The authors have no relevant financial or non-financial interests to disclose.

## Authors contributions
***Abdelghani Maddi***: *conceptualization; data curation; formal analysis; investigation; methodology; validation; visualization; writing – original draft preparation; writing – review and editing.*
***Emmanuel Monneau***: *conceptualization; data curation; methodology; formal analysis; writing – original draft preparation; writing – review and editing.*
***Catherine Guaspare-Cartron***: *conceptualization; formal analysis; writing – original draft preparation; writing – review and editing.*
***Floriana Gargiulo***: *conceptualization; writing – review and editing.*
***Michel Dubois***: *project administration; supervision; conceptualization; formal analysis; writing – original draft preparation; writing – review and editing.*

## Funding information
This work was supported by a grant overseen by the French National Research Agency (ANR). Grant number: ANR-20-CE26-0008. Website: https://anr.fr/Projet-ANR-20-CE26-0008.

## 1. Introduction

The launch of PubPeer in 2012 represented a significant milestone in the promotion of Post-Publication Peer Review (PPPR). This platform played a crucial role in advancing PPPR, primarily owing to the unique feature it introduced— the possibility of commenting on published articles with complete anonymity (Teixeira da Silva 2018). In contrast to other platforms like PubMed Commons that asked users to identify themselves, PubPeer provided a safe space for researchers to express their opinions. This significant aspect has fostered an online environment free from concerns of potential professional repercussions to facilitate the identification of errors, the reporting of issues in articles, and the detection of instances of scientific misconduct (Blatt 2015). Importantly, this approach broadened the scope of PPPR by encouraging a more diverse range of researchers to actively participate in the evaluation of scientific research, further solidifying PubPeer's place as a transformative force in scholarly discourse.

The literature on PubPeer, reflects a series of studies that have analyzed the platform since its inception. In their chapter, (Barbour et Stell 2020), the founders of PubPeer, presented the purpose of the PubPeer platform and staunchly defended the concept of anonymity and its pivotal role in the development of PPPR. They unequivocally concluded that PPPR represents the future of research evaluation, advocating for its potential to complement, or even supplant, the current quantitative indicator-based evaluation system. However, the discussion surrounding PubPeer's transparency extends beyond Barbour and Stell's perspective. Authors like (Teixeira da Silva 2018), who scrutinized the platform's "about" page, highlighted the need for comprehensive transparency in all aspects of PubPeer's operations to enhance its credibility as a PPPR platform. Texiera's study concluded by emphasizing that organizations tasked with overseeing the scientific community must, in turn, be subject to scrutiny themselves.

Several other studies have delved into specific aspects of the PubPeer platform. In 2019, Dubois and Guaspare studied the discussions on Pubpeer devoted to the publications of Olivier Voinnet, a French biologist suspected of scientific misconduct (Dubois et Guaspare 2019). They showed that PubPeer contributors, far from freeing themselves from the norms of the scientific community, produce new lines of demarcation between what is considered acceptable or not in a research community, and more globally what distinguishes good research practice from scientific misconduct. In 2020, (Bordignon 2020) conducted an analysis to determine whether negative citations in articles and comments posted on PPPR platforms equally contribute to the correction of scientific literature. Her research revealed that comments on PubPeer contribute more significantly to self-correction of science than negative citations do. Additionally, (Ortega 2021)'s study examined the relationship between errata, expressions of concern, and retractions within a sample of 39,449 publications commented upon in PubPeer, finding a relatively weak correlation among them. Building upon this, (Ortega 2022)'s work classified PubPeer comments based on their severity, ranging from relatively rare positive comments to those exposing fraud or manipulation. In a more recent study by (Ortega et Lorena Delgado-Quirós 2023), they analyzed editorial responses to comments posted on PubPeer. Out of a sample of 17,244 commented publications, only 21.5% received feedback from editors, highlighting a need for improved engagement between the platform and the academic community. These diverse studies shed light on various facets of PubPeer's role and functionality within the scientific discourse.

The present article contributes to the growing literature on the prominent PPPR platform, PubPeer, by introducing a novel analysis that, for the first time, distinguishes the gender of authors of commented publications and examines collaborations between male and female researchers in this context. The primary objective of this study is to analyze the fate of publications subjected to commentary on PubPeer while concurrently scrutinizing the gendered aspect of these publications. This unique approach aims to enhance our understanding of PPPR dynamics while shedding light on issues related to gender diversity in scientific research.

This study is structured around two main research questions: 1) Among commented publications on PubPeer, does the collaboration between male and female authors influence the (non)retraction of scientific publications? 2) For publications that have both received comments on PubPeer and been retracted, are there any disparities in the reasons for retractions based on the collaboration between men and women?

By addressing these questions, we hope to contribute valuable insights to the broader discussion on gender-related factors in scientific publishing and the potential impact of public commentary on the validity and retraction of publications. Gender disparities and issues of correction of science have garnered increasing attention in the scientific literature. Nevertheless, our understanding of these topics, particularly regarding the influence of gender collaboration on publication retractions, remains limited.

The subsequent sections of this article are organized as follows: Sections 2 and 3 provide a review of the literature on women in science and the gendered self-correction of science, respectively. In Section 4, we outline the research questions, data and methodology utilized in our study. Section 5 presents the detailed results obtained from our analysis. Finally, in Section 6, we offer a discussion of the findings, their implications, and potential avenues for further research and potential recommendations.

## 2. Collaboration in Science and Gender Differences

Throughout the twentieth century, the presence of women in science grew steadily. Although the number of women has increased in all scientific disciplines (Rossiter 1997, 2012), numerous studies in the sociology of science have highlighted the persistence of gender disparities across many dimensions of the academic career (salaries, career achievement, prestigious status or awards…) (Larivière et al. 2013; Lincoln et al. 2012; Rossiter 1993). More, these disparities between men and women age peers widen between the end and the beginning of their careers (Sugimoto et Larivière 2023; Zuckerman, Cole, et Bruer 1991).

Since the 1980s, quantitative analyses have repeatedly shown that women's scientific productivity is generally lower than that of men (Cole et Zuckerman 1984; Reskin 1978; Tower, Plummer, et Ridgewell 2007). To quote some recent results, except a few highly feminized disciplines (arts and health) where the gap between men's and women's publications is small, men publish 20% more than women across all disciplines, with the gap reaching a third (mathematics, physics, social sciences) and more (50% in psychology) (Sugimoto et Larivière 2023). This trend of studies has especially emphasis the link between gender gap in access to a range of gains in academia and gender disparities in scientific production and citation insofar as authorship plays a key role in access to scientific advantages (position, status, funding, prizes…).

In their last enlightening work, Sugimoto and Larivière (2023), relying on scientometric analyses of millions of publications in all disciplines, make an essential contribution to the comprehension of the mechanisms that hinder the development of women's career, and that are largely rooted in the functioning of the authorship attribution. Paying attention to the distribution of authorship in publications, they highlight a systematic disadvantage of women, who are less likely to occupy positions of value in bylines. Among a range of eye-opening analyses, they make strong evidence that women are less likely than men to write as a sole author, whereas signing alone is particularly rewarding. If single-authored articles have largely decreased with the rise of collective authorship in all disciplines since the Second World War, single authorship reflects a type of intellectual labor considered of high scientific value not to mention the high citation rate sole authored articles usually capture.

If they publish less as single author, they are more likely than men to write in collaboration. But in all fields, it is men who occupy the dominant and therefore rewarding positions in the byline, i.e. first and last author. Women tend to occupy more often positions in the byline that mark a junior status or a low position in the career, or that indicate the performance of technical tasks of relatively low prestige (i.e. early and mid-positions). However, they are more likely to occupy a leadership role by signing as first author if the size of the team increases. Moreover, above a certain team size, they tend to lose this dominant position, a trend reflected in the analyses based on our data.

**3. Correction of science and gender differences**

Numerous studies have delved into the realm of retracted scientific articles, exploring the multifaceted aspects surrounding them (Fang et Casadevall 2011; Wray et Andersen 2018). These investigations have examined diverse angles, encompassing the reasons for retractions, ethical considerations, methodological flaws, and the potential consequences for scientific trust and credibility (Cokol, Ozbay, et Rodriguez-Esteban 2008; Pfeifer et Snodgrass 1990; Steen 2011; Steen, Casadevall, et Fang 2016). Research in this area has not only deepened our understanding of the complex landscape of retracted articles but has also underscored their importance in the broader context of scientific integrity and scholarly communication.

Gender disparities in science extend beyond representation and career advancement; they also manifest themselves in areas such as the correction and retraction of scientific work. Existing literature has shown that there are notable gender differences in acknowledging errors, recanting and correcting scientific findings. Studies have shown that female scientists are more likely to backtrack on their research or admit mistakes than their male counterparts (Wadman 2005). This gender disparity in correctional practices can be attributed to a variety of factors including societal expectations, biases and cultural norms that put additional pressure on women to uphold high standards of accuracy and integrity (Kaatz, Vogelman, et Carnes 2013). Additionally, systemic biases and gender inequalities within the scientific community can influence the reception and response to errors, leading to differential treatment based on gender. Gender differences in scientific misconduct is a topic that has gained increasing attention in the recent scientific literature, with several studies shedding light on various aspects of this issue. (Fang, Bennett, et Casadevall 2013) conducted a comprehensive analysis of the Office of Research Integrity (ORI) database, examining 228 cases of scientific misconduct. Their findings revealed that a significant majority of these cases involved male researchers, particularly professors. However, it is important to note that these findings do not imply that male researchers are inherently less ethical than their female counterparts. (Kaatz et al. 2013) further emphasized the need to investigate the historical and sociological factors contributing to these disparities.

Examining the presence of women in retracted publications, (Decullier et Maisonneuve 2021) conducted one of the first studies on this topic. Their analysis of 113 retracted publications revealed that 37.2% of them were authored by women, either as the first author or as sole authors. Notably, fraud and plagiarism accounted for 28.6% and 59.2% of retractions, respectively, for both men and women. In the field of biomedical sciences, (Pinho-Gomes, Hockham, et Woodward 2023) examined gender differences in the authorship of retracted articles, utilizing data from Retraction Watch. Among 35,635 retracted biomedical articles spanning from 1970 to 2022, women accounted for 27.4% of first authors and 23.5% of last authors. The underrepresentation of women was particularly pronounced in cases of fraud (approximately 19%). Notably, the highest representation of women as first authors was observed in cases related to editorial issues (35%) and errors (29.5%). Overall, the majority of retractions (60.9%) involved male researchers as both first and last authors. The authors suggested that achieving gender equality could contribute to enhancing research integrity in the field of biomedical sciences.

In a study by (Ribeiro et al. 2023), the authors investigated the behavior of self-retraction due to honest errors, considering factors such as country, research domain, and gender. Analyzing 2,906 retracted publications categorized under "errors" between 2010 and 2021, the study revealed that women scientists accounted for 25% of self-retractions due to honest errors. However, further research is warranted to develop standardized indicators that measure the proportional representation of women in science, specifically in the context of retractions. Furthermore, (Satalkar et Shaw 2019) conducted interviews with 33 Swiss researchers to explore the early influences on concepts of honesty, integrity, and fairness in research. The findings highlighted the significance of early education, moral values instilled by families, and participation in team sports as the primary influencers. Notably, two-thirds of the participants had not received any formal training on research integrity, which may increase the likelihood of honest errors due to a lack of knowledge.

## 4. Research Questions, Data and Method

### 4.1. Research Questions

In this research, we delve into the domain of self-correction of science within the PubPeer database, seeking to shed light on the influence of male-female collaborations on the (non)retraction of scientific publications. The literature has demonstrated that male-female collaborations act as catalysts for generating more impactful and groundbreaking scientific research (Maddi et Gingras 2021). Still, little is known about how such collaborations might impact the correction process and the subsequent retraction rates of published works.

Under the prism of male-female collaborations, we embark on a comprehensive analysis of the PubPeer database, scrutinizing retracted publications and their reasons, aiming to discern potential disparities between male-female collaborative dynamics. Our investigation seeks to unravel whether gender-based collaborations are associated with distinct patterns in the reasons for publication retractions. Such insights could contribute to a more nuanced understanding of scientific integrity and the factors that underpin the reliability of research outcomes.

This study not only holds the potential to uncover critical nuances in the scientific correction process but also contributes to the broader discourse on gender representation in scientific collaborations and their impact on the scientific literature. By illuminating these aspects, we aim to foster a more robust and equitable scientific ecosystem that bolsters the credibility and reproducibility of research findings.

To achieve these objectives, we articulate two pivotal research questions guiding our exploration:

**RQ 1:** *"Does male-female collaboration significantly influence the (non)retraction rates of scientific publications, and if so, to what extent does it transform the overall retraction dynamics?"*

**RQ 2:** *"Are there discernible disparities in the reasons underlying the retraction of scientific publications, depending on the collaborative dynamics between male and female researchers?"*

*4.2. Publication's data*

Our publication data is extracted from the PubPeer database, which includes publications that have been commented between 2012 (the launch date of the database) and 2021. We enriched the dataset with metadata sourced from OpenAlex (providing publication openness status) and Web of Science (for disciplinary information).

From the *PubPeer* website, we have built data on this platform about three main entries: comments, commentators and commented publications.

As far as publications are concerned, the information available on *PubPeer* (in particular the title and the URL) has been supplemented by a set of information from the *OpenAlex API*. Thus, when possible, each publication was provided with identifiers (PubMed, OpenAlex, mag, DOI, ISSN), a title, a source, a date, statutes concerning open access and retraction, key-words, an abstract, Open Alex concepts, authors, authors' institutions, countries of affiliation of authors' institutions, a number of citations. Using the ISSN identifier, discipline categorizations were associated with these publications with *Web of Science* and *Scimago*.

The data on *PubPeer* that we have built covers a 10-year period, from its creation in January 12, 2012 to December 31, 2021. There were 189,426 comments, 42,661 commentators and 101,272 commented publications over this decade. In this study, we excluded from the analysis the publications whose gender was not identified for any author, i.e. 2113 publications. This reduces the number to 99159 publications analyzed for global statistics. In analyzes involving the identification of the first and or last author, we also excluded publications for which the sex of the authors of these positions was not identified (for example in Table 2). Similarly, for the regression, we excluded publications that are not indexed in the Web of Science as far as we used the discipline. In summary, we adapted our sample according to the object analyzed.

*4.3. Retraction data*

The data on retracted publications were kindly provided by the RetractionWatch (https://retractionwatch.com/) database provider in April 2023. The database contains nearly 40,000 retracted publications. For this study, we matched the retraction data with the PubPeer database using the DOI. In addition to information regarding the retraction status of publications, we also utilized the reasons for retraction to analyze gender differences. To accomplish this, we identified all retraction reasons and aggregated them as presented in Appendix 1[1]. To avoid arbitrary choices in retraction reasons, we employed a fractional counting approach, considering all reasons. For instance, if a publication was retracted for two reasons, A and B, we assigned a count of 0.5 to each reason. This approach has two advantages: firstly, it allows us to measure the relative intensity of each retraction reason by type of collaboration; secondly, it enables us to sum the counts so that the total count for all reasons and collaboration types equals the total number of retractions (i.e., 4410). In this analysis, we focus on the top 20 reasons in terms of frequency. Appendix 2[2] provides the complete distribution of reasons based on their frequency, by type of men-women collaboration, in our sample.

*4.4. Gender data*

To identify the gender of authors, several methods exist (Larivière et al. 2013; Wais, Kamil 2016; West et al. 2013). In this paper, we opted to utilize (Wais, Kamil 2016)' method and the R package "genderize.io" (see Appendix 3[3] for more details).

---

[1] Appendix 1 : https://figshare.com/s/f70e9c14b9c2b6beb2c0
[2] Appendix 2 : https://figshare.com/s/2c090bba3ed75b89a666
[3] Appendix 3 : https://figshare.com/s/fc0b93bb1441068576c1

**Table 1. Distribution of publications commented on PubPeer by research area and authorship**

| Research area | # (%) of male authors | # (%) of female authors | Average proportion of women by paper |
|---|---|---|---|
| **Physical Sciences**, N = 18,022 | 13,362 (74%) | 4,660 (26%) | 27,0% |
| **Technology**, N = 21,281 | 15,455 (73%) | 5,826 (27%) | 28,5% |
| **Multidisciplinary**, N = 37,553 | 25,044 (67%) | 12,509 (33%) | 34,1% |
| **Life Sciences Biomedicine**, N = 242,541 | 154,654 (64%) | 87,887 (36%) | 37,0% |
| **Arts Humanities**, N = 1,293 | 769 (59%) | 524 (41%) | 40,7% |
| **Social Sciences**, N = 133,235 | 74,100 (56%) | 59,135 (44%) | 44,4% |
| *Overall, N = 453,925* | *283,384 (62%)* | *170,541 (38%)* | *38,4%* |

Table 1 gives an overview of publications commented on PubPeer by research area and authorship. It reveals that, for example, of the 18,022 authors listed in publications in the physical sciences, 26% of them are women. These data highlight a significant gender disparity within the field. Moreover, when we consider the contribution of women in these publications, measured by the average proportion of women per article, we find that women hold a weight of 27%.

The table also highlights the phenomenon of horizontal segregation in research, which refers to the concentration of women in certain disciplines while they are under-represented in others. In this case, the data reveals a lower presence of women in applied disciplines and a stronger presence in the fields of life sciences and biomedicine. This observation aligns with the well-known trend of gender disparities in STEM fields, where women tend to gravitate toward certain fields while being underrepresented in others.

The relatively high proportion of women (38%) in the PubPeer data compared to their presence in the global data, 33% (UNESCO 2015), can be attributed to the significant presence of social sciences (especially psychology), life sciences and biomedicine publications in the dataset. These fields, which have a strong female presence, receive substantial attention and commentary in PubPeer.

**Figure 1: Distribution of publications based on whether the main author is a woman or a man, according to the team size (number of authors)**

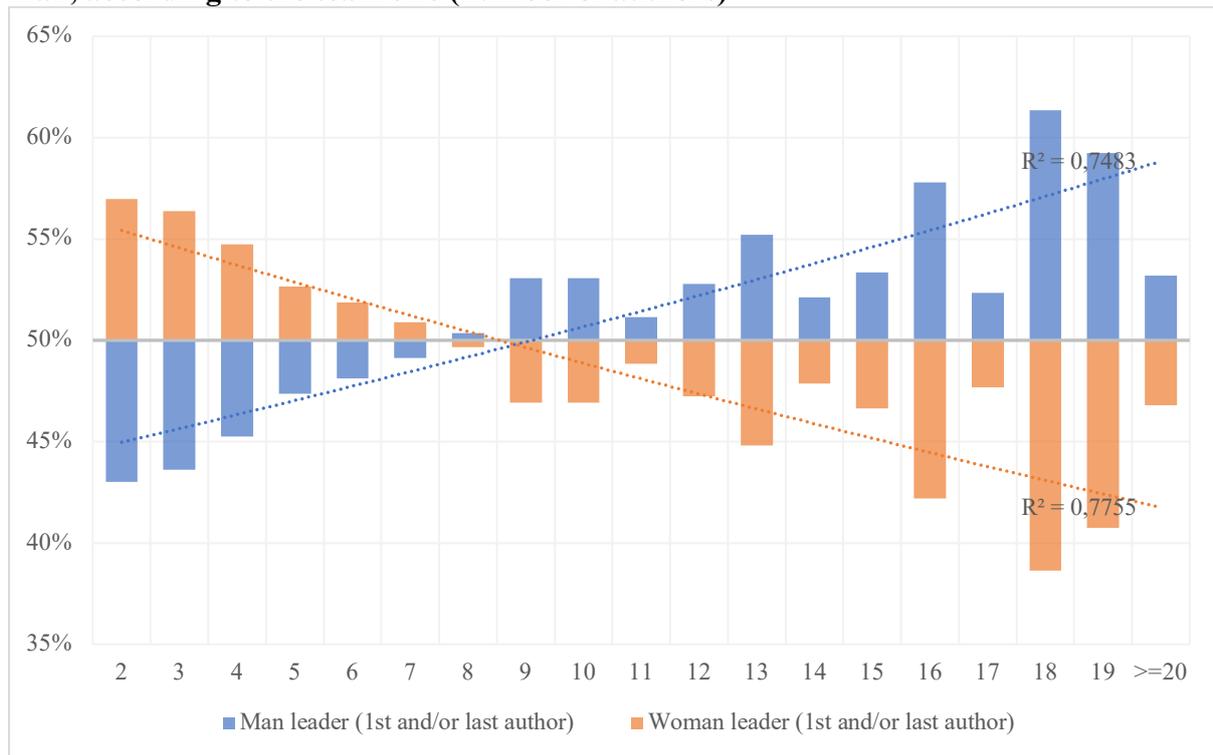

Figure 1 shows the distribution of publications based on whether the main author is a woman or a man, according to the team size (number of authors). As it can be seen in this figure, in small teams (with less than 8 authors), women tend to be more prevalent as leaders, assuming the role of the main author (first author and/or last author). However, for larger teams, it is predominantly men who take up the leadership role as the main authors. This finding is consistent with existing literature in scientometrics and gender studies within academia.

Numerous studies in scientometrics have explored the patterns of authorship and publication productivity based on gender. One possible explanation for the observed phenomenon is the so-called "Matthew effect" or "accumulative advantage" hypothesis. According to this theory, researchers who have already achieved success and recognition (typically more experienced male researchers) tend to join larger research collaborations, which further consolidates their status and provides them with more opportunities to lead and secure high-profile projects. This accumulation of advantages may explain why men are more likely to be main authors in larger teams.

On the other hand, for small teams, it is possible that factors like collaborative dynamics and equal distribution of responsibilities play a more significant role in determining the main author. In such settings, women may have a better chance of taking the lead, as collaborative efforts are often more egalitarian, and individual contributions may be more readily acknowledged and valued.

Additionally, the "Matilda effect" also deserves consideration in this context. The Matilda effect refers to the systematic undervaluation and marginalization of the contributions of women in science and academia throughout history. This phenomenon suggests that women's achievements and discoveries are often attributed to their male colleagues or overlooked entirely, leading to a lack of recognition and opportunities for women researchers.

*4.5. Regression analysis*

We conducted a logistic regression analysis, with the binary outcome variable indicating whether a publication was retracted or not (coded as 1 for retracted). The explanatory variables consisted of indicators representing different types of male-female collaborations, using publications involving only women (more than one woman) as the reference category. We controlled for the number of authors, publication openness status, and discipline. To analyze the role of control variables, we progressively built three models: first incorporating the number of authors and whether the publications were open access or not, followed by the discipline.

## 5. Results

In this section, we present the main results of our analysis. Firstly, we examine the distribution of publications based on the type of men-women collaboration in the overall PubPeer dataset and within the retracted publications. Secondly, we present the distribution based on the retraction reasons. Finally, we discuss the results of the logistic regression.

*4.1. Men-women collaboration and retracted publications*

Table 3 displays the distribution of commented publications in PubPeer according to the type of men-women collaboration, with a distinction between retracted and non-retracted publications. We classify publications into three main types:

- Publications involving collaboration between men and women, distinguishing whether the principal author is a man or a woman.
- Publications involving only men or only women.
- Publications with a single author (either male or female).

Table 2 reveals that the majority of publications result from H-F collaboration, with 34% having a male corresponding author and 31% having a female corresponding author. In contrast, there is a significant disparity within publications involving single-sex collaboration: publications involving only men account for 21%, while those involving only women represent only 7.8%. Single-author publications are the least prevalent, comprising 4.3% for men and 2% for women.

**Table 2: distribution of publications according to the type of men-women collaboration in the overall PubPeer dataset and within the retracted publications**

|  | Overall, N = 93,563 | Is retracted |  |
|---|---|---|---|
|  |  | No, N = 89,050 | Yes, N = 4,513 |
| **Gender collaboration type** |  |  |  |
| Collab. M-W (Man principal author) | 31,468 (34%) | 29,591 (33%) | 1,877 (42%) |
| Collab. M-W (Woman principal author) | 28,853 (31%) | 27,624 (31%) | 1,229 (27%) |
| Collab. men only | 20,005 (21%) | 19,039 (21%) | 966 (21%) |
| Collab. women only | 7,305 (7.8%) | 7,157 (8.0%) | 148 (3.3%) |
| Man alone | 4,052 (4.3%) | 3,839 (4.3%) | 213 (4.7%) |
| Woman alone | 1,880 (2.0%) | 1,800 (2.0%) | 80 (1.8%) |

Among the retracted publications, while the proportion of publications involving only men remains at 21%, other types of publications exhibit notable differences. Specifically, publications involving men-women collaboration with a male corresponding author represent 42% of the retracted publications, which is substantially higher than the 33% observed in the overall commented publications. Conversely, publications involving only women account for 3.3% of the retracted publications, compared to their representation of over 7.8% in the total commented publications.

To facilitate analysis, Figure 2 presents a double ratio that compares the proportion of each publication type within the retracted publications to its corresponding proportion in the overall dataset. We observe that publications involving men-women collaboration with a male corresponding author are 24% more prevalent (ratio of 1.24) in the corpus of retracted publications than in the total dataset. In contrast, publications involving only women are 58% less prevalent (ratio of 0.42) in the retracted publications compared to the overall commented publications.

**Figure 2: Proportion in retracted publications divided by proportion overall, by male female collaboration**

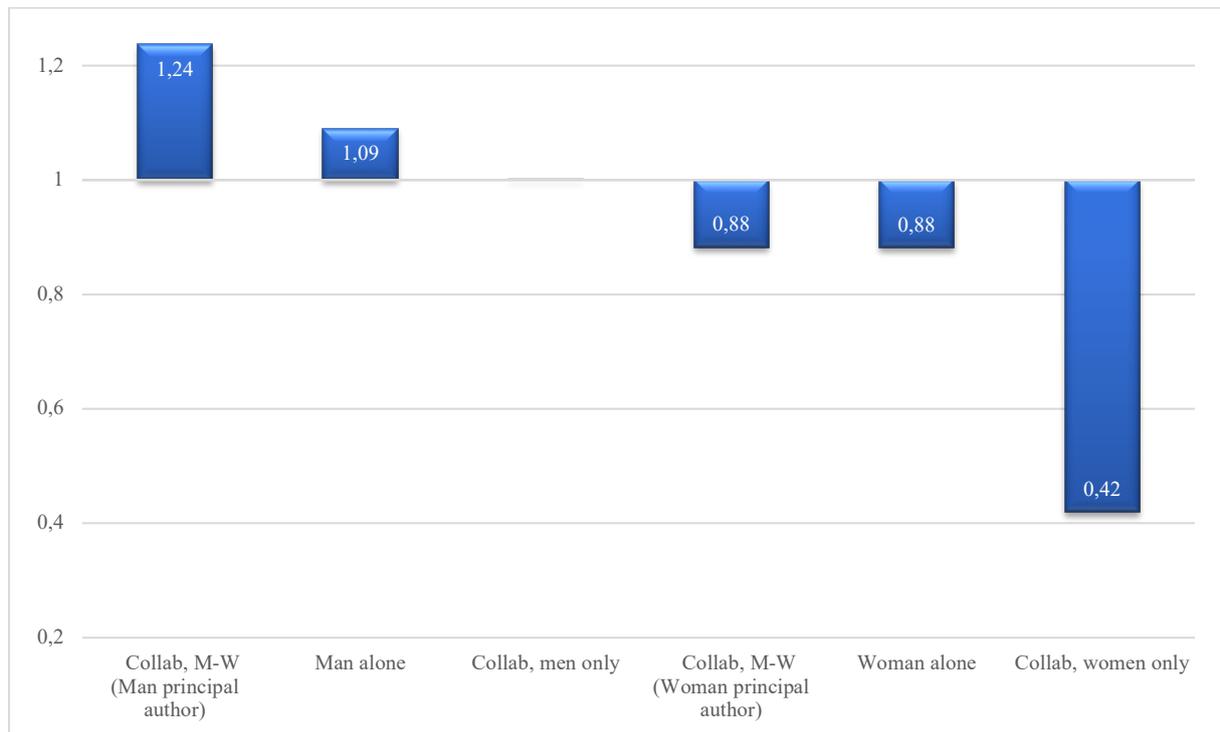

The results from Table 2 and Figure 2 provide insights into the structure of retracted publications based on gender and H-F collaboration. Publications led by men are generally more prone to retraction. However, these findings require further confirmation and controlling for certain variables such as discipline, total number of authors per discipline, and the open access status of publications. This analysis is carried out in the subsequent section on logistic regression.

*4.2. Reasons of retraction*

Table 3 presents the distribution of the 4,513 retracted publications according to the reasons for retraction and the type of men-women collaboration. As mentioned in the Methods section, we employed a fractional counting approach. Only the top twenty reasons are displayed (for the complete list, see Appendix 2). It is observed that within this sample of retracted publications among the commented publications in PubPeer, "Duplication of Image" is the most frequent reason, followed by "Concerns/Issues About Data." This finding is noteworthy as image-related comments constitute a significant portion and are widely practiced on this platform. The table also highlights disparities in the reasons for retraction based on the type of H-F collaboration. However, in order to thoroughly analyze these results, it is important to correct for the effect of sample size.

**Table 3: Distribution of reasons for withdrawal by type of collaboration men women, fractional accounting**

| | Man as leader of publication | | | Woman as leader of publication | | | |
|---|---|---|---|---|---|---|---|
| Reasons (Retraction Watch) | Collab. men only | Man alone | Collab. M-W (Man principal author) | Collab. women only | Woman alone | Collab. M-W (Woman principal author) | Total |
| Duplication of Image | 92 | 12 | 264 | 19 | 4 | 181 | 571 |
| Concerns/Issues About Data | 71 | 13 | 144 | 13 | 4 | 91 | 336 |
| Unreliable Results | 49 | 6 | 107 | 9 | 2 | 63 | 236 |
| Investigation by Third Party | 48 | 11 | 89 | 7 | 5 | 50 | 209 |
| Paper Mill | 43 | 9 | 88 | 8 | 5 | 50 | 203 |
| Manipulation of Images | 29 | 3 | 81 | 5 | 1 | 70 | 189 |
| Investigation by Journal/Publisher | 43 | 16 | 60 | 5 | 7 | 45 | 177 |
| Original Data not Provided | 33 | 3 | 70 | 3 | 1 | 50 | 161 |
| Concerns/Issues About Image | 29 | 5 | 75 | 6 | 1 | 44 | 161 |
| Investigation by Company/Institution | 36 | 6 | 63 | 5 | 1 | 48 | 159 |
| Duplication of Article | 51 | 13 | 39 | 2 | | 17 | 121 |
| Upgrade/Update of Prior Notice | 19 | 6 | 42 | 3 | 5 | 27 | 100 |
| Error in Image | 14 | 1 | 45 | 1 | 1 | 37 | 98 |
| Unreliable Data | 17 | 4 | 42 | 2 | | 24 | 89 |
| Plagiarism of Article | 22 | 12 | 28 | 6 | 4 | 17 | 87 |
| Author Unresponsive | 20 | 5 | 32 | 2 | 3 | 19 | 81 |
| Error in Data | 12 | 3 | 36 | 1 | 1 | 27 | 79 |
| Notice - Limited or No Information | 17 | 7 | 28 | 5 | 2 | 19 | 78 |
| Fake Peer Review | 24 | 9 | 23 | 2 | 5 | 13 | 76 |
| Results Not Reproducible | 12 | 1 | 38 | 2 | 1 | 19 | 72 |
| **Total top 20 reasons** | 681 | 143 | 1394 | 104 | 50 | 911 | 3283 |
| **Total other reasons** | 285 | 70 | 483 | 44 | 30 | 318 | 1230 |
| **Total** | 966 | 213 | 1877 | 148 | 80 | 1229 | 4513 |

To measure the relative weight of different reasons according to the type of men-women collaboration, we applied the method of calculating the activity index traditionally used in bibliometrics to measure disciplinary specialization. In our case, this indicator is calculated as follows. For example, for the reason "Duplication of Image" and collaborations involving only men, the indicator is obtained by comparing the proportion of "Duplication of Image" in the total retracted publications involving only men to the proportion of "Duplication of Image" in the overall dataset. The indicator varies around 1. Thus, for the given example, the corresponding value is 0.79, indicating that the reason "Duplication of Image" is 21% less prevalent in retracted publications involving only men compared to the overall dataset.

**Figure 3: Ratio between the share of each reason by type of collaboration on the same share on the whole basis of retractions**

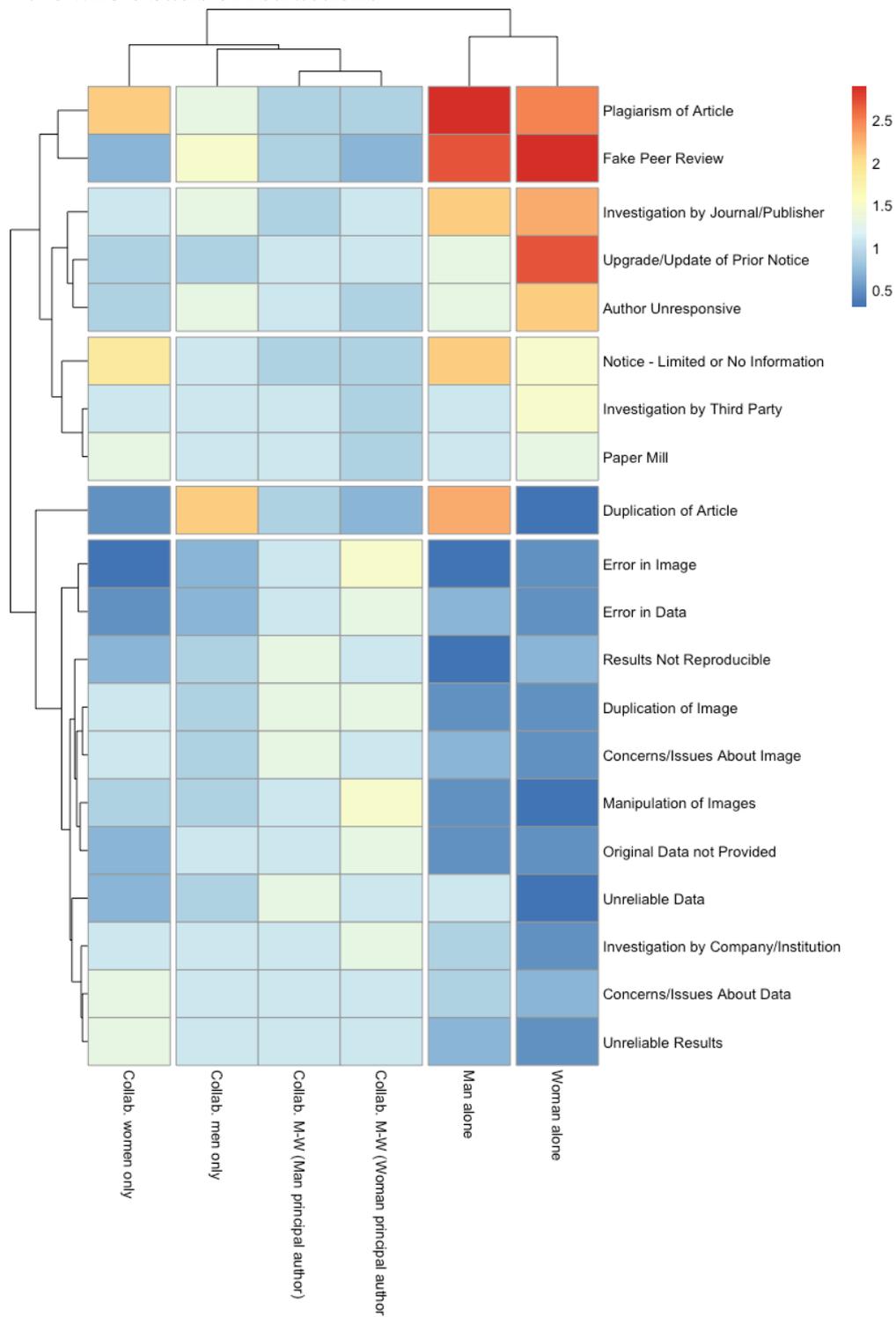

Figure 3 highlights several interesting findings. First, it is worth noting that there are similarities between men and women when they are the sole authors of a publication, which is sociologically interesting. For instance, the results suggest that when men publish articles alone, they are more likely to engage in "Fake Peer Review," a practice involving recommending a fake reviewer during the peer review process, either by proposing existing scientists while providing false email addresses or by suggesting fictional reviewers completed with an email address (Bouter et al. 2016; Ferguson, Marcus, et Oransky 2014). It is noteworthy that "Fake Peer Review" is highly prevalent in retractions involving collaborations between men only (45% higher than the overall dataset). Another characteristic of retracted publications involving a single author (male or female) is the significant presence of retractions due to "Plagiarism of Article". This reason is also prominent for collaborations between men only, albeit to a lesser extent.

Another striking result is the strong concentration of publications with a female lead author (regardless of collaboration) in image-related issues such as duplication, errors, or manipulation. "Error in Data" is also frequently observed when the lead author is a woman. In publications involving only men (either alone or in collaboration), the reason "Duplication of Article" is highly prevalent compared to their female counterparts. Finally, for collaborations involving only women, the main reason is "Concerns/Issues About Data," which is 35% higher than the average.

However, it is important to note that the disciplinary distribution of reasons for retraction could have an impact on the results obtained, as the distribution of women and men differs across disciplines.

*4.3. Regression analysis*

Table 4 presents the results of the logistic regression. The dependent variable is the retraction status of a publication. Due to data availability limitations, particularly in terms of disciplines, the sample was reduced to 77,608 publications, including 3,993 retractions. We proceeded iteratively by including a block of variables in each model. In Model 1, only gender-related variables were integrated. In subsequent models, several control variables were progressively included: the number of authors per publication (Model 2), the open access status (Model 3), and the discipline (Model 4). As we can see the model's goodness ($R^2$ and AIC/BIC) of fit improves as new variables are incorporated.

**Table 4: Logistic Regression Results: Coefficients and p-values**

| Variable type | Explicative variables | Model 1 | | Model 2 | | Model 3 | | Model 4 | |
|---|---|---|---|---|---|---|---|---|---|
| | | Coeff. | p.value | Coeff. | p.value | Coeff. | p.value | Coeff. | p.value |
| **Intercept** | Intercept | -3,87 *** | 0,00E+00 | -4,41 *** | 0,00E+00 | -4,5 *** | 0,00E+00 | -2,17 *** | 1,15E-107 |
| **M-W collaboration** | Collab. M-W (Man principal author) | 1,11 *** | 4,15E-35 | 0,71 *** | 1,18E-16 | 0,68 *** | 4,25E-13 | 0,29 *** | 3,10E-03 |
| | Man alone | 0,99 *** | 8,85E-18 | 1,54 *** | 2,29E-39 | 1,34 *** | 6,92E-29 | 0,23 * | 6,20E-02 |
| | Collab. men only | 0,90 *** | 4,68E-22 | 0,82 *** | 1,73E-20 | 0,79 *** | 4,60E-17 | 0,20 ** | 3,59E-02 |
| | Woman alone | 0,55 *** | 4,68E-04 | 1,1 *** | 1,58E-14 | 0,96 *** | 2,78E-09 | 0,03 | 8,46E-01 |
| | Collab. M-W (Woman principal author) | 0,78 *** | 2,71E-17 | 0,43 *** | 5,71E-09 | 0,38 *** | 4,77E-05 | 0,02 | 8,06E-01 |
| | *Collab. women only (ref)* | - | - | - | - | - | - | - | - |
| **Scientific collaboration** | Number of authors (log transformed) | - | - | 0,54 *** | 1,01E-77 | 0,33 *** | 4,92E-27 | -0,28 *** | 6,37E-18 |
| **Openness** | Is open access | - | - | - | - | 0,84 *** | 6,07E-125 | 0,33 *** | 7,15E-20 |
| **Discipline** | Social Sciences | - | - | - | - | - | - | -3,7 *** | 1,44E-301 |
| | Physical Sciences | - | - | - | - | - | - | -0,27 *** | 1,09E-04 |
| | Technology | - | - | - | - | - | - | -0,07 | 2,29E-01 |
| | Arts Humanities | - | - | - | - | - | - | -0,35 | 5,07E-01 |
| | *Life Sciences Biomedicine (ref)* | - | - | - | - | - | - | - | - |
| **Model statistics** | R^2 | 0,007 | | 0,018 | | 0,036 | | 0,151 | |
| | AIC | 32436.25 | | 32098.87 | | 31515.06 | | 27751.64 | |
| | BIC | 32492.28 | | 32164.24 | | 31589.77 | | 27863.70 | |

Dependent variable: if the commented publication is retracted. Yes = 3993 (4.9%), No = 77608 (95.1%)

(*): significant at 10%, (**): significant at 5%, (***): significant at 1%.

An initial interesting finding is that when control variables are not included (Model 1), we obtain exactly the same order in terms of the impact of different types of men-women collaborations as observed in Figure 2. "Collab. M-W (Man principal author)" has the highest coefficient, followed by "Man alone," and so on. It is worth noting that all coefficients are positive and statistically significantly different from zero at a risk level below 1%. Once the "number of authors" variable is introduced as a control variable (Model 2), the order changes significantly. The coefficient for this variable is positive and significantly different from zero, as is the coefficient for the open access status variable (Model 3).

The results obtained using the equation of Model 4 (which incorporates disciplines) are striking. Firstly, it is evident that only variables characterizing publications led by men are statistically significant. In other words, these publications are 20% to 29% more likely to be retracted than those involving only women. The coefficients for these variables are also higher than those for "Woman alone" and "Collab. M-W (Woman principal author)" variables. Furthermore, the results indicate no statistically significant difference between the various types of publications with a female lead. The distinction is thus made between publications led by men versus those led by women.

Secondly, it is observed that while the coefficient for the "number of authors" variable remains highly significant, it becomes negative once disciplines are included in the equation. This can be explained by the fact that the positive impact observed in previous models is driven by disciplines with both a high proportion of retractions and a large number of authors. Once the disciplinary effect is controlled for, the number of authors per paper negatively impacts retractions. This result is interesting for several reasons. Firstly, it suggests that a large number of authors per publication may indicate greater control, and thus fewer errors. Additionally, one could hypothesize that authors are less inclined (or less willing) to engage in questionable practices when collaborating with colleagues.

Finally, a final finding is that open access publications are generally more retracted than their subscription-based counterparts. However, due to the nature of our sample consisting of journals indexed in WoS, we cannot draw conclusions about the prevalence of journals with a lax peer-review process. Nevertheless, certain publishers indexed in this database, such as MDPI, are categorized as "grey" publishers with questionable practices, to the extent that WoS has delisted some of their journals. Therefore, it would be valuable to further investigate this aspect in a future study.

## 6. Discussion and conclusion

Our study is a contribution to the emerging field of studies devoted to post-publication peer review platforms. Its aim is twofold. Firstly, it aimed to investigate the disparities in collaboration patterns between men and women and their association with retracted publications, specifically within the discussions on PubPeer. Secondly, we sought to analyze the variations in retraction reasons based on the gender composition of research teams. To accomplish our objectives, we employed a comprehensive dataset that integrated information from PubPeer, Retraction Watch, Web of Science, and OpenAlex, resulting in a sample size exceeding 90,000 publications.

One of the most notable findings of our analysis is the pronounced disparity in retraction rates between publications led by men and those led by women within the PubPeer platform. Publications led by men were found to experience a higher incidence of retractions compared to those led by women.

Another significant outcome of our study is the identification of gender-based discrepancies in retraction reasons. Notably, women exhibited a relatively higher likelihood of retraction due to issues related to image manipulation, duplication, or errors. On the other hand, men were more prone to retractions caused by article duplication. Furthermore, single-authored publications, irrespective of the author's gender, displayed an increased risk of retraction stemming from fake peer reviews.

Furthermore, our investigation revealed an intriguing trend regarding the higher retraction rates among open access publications compared to their subscription-based counterparts. Future research is warranted to examine whether this observation can be attributed to the practices of "gray" publishers indexed in international databases such as MDPI or Hindawi (Nicholas et al. 2023).

While our findings may appear to reinforce societal stereotypes regarding men, it is crucial to approach the issue with a more nuanced understanding and delve deeper into the intricate dynamics at play. For instance, the unequal distribution of research funding may contribute to a higher likelihood of men engaging in fraudulent behavior. Moreover, individual behavior can be influenced by various factors, including ambition and career aspirations, which often subject men to heightened pressures. Consequently, when faced with a disparity between their ambitions and actual capabilities, some men may resort to harassment as a means to attain their desired outcomes. Notably, this behavior tends to intensify in the presence of female competition. In contrast, women tend to perceive themselves as more inclined towards ethical conduct, potentially influenced by the social desirability bias—a tendency to present oneself in a favorable light—which affects women to a greater extent (Decullier et Maisonneuve 2021; Decullier et Sèdes 2022).

The observations made in our study prompt us to delve deeper into the questions initially raised (Fang et al. 2013). Specifically, we must consider: What motivates individuals to commit research misconduct? Does competition for prestige and resources disproportionately drive misconduct among male scientists? Are women more sensitive to the threat of sanctions? Is gender a correlate of integrity? These are essential questions that warrant further analysis and exploration. Understanding the complex interplay of these factors is crucial in comprehending the observed disparities in publication retractions.

In summary, our research sheds light on discrepancies in collaboration trends and retractions in scholarly publications. This highlights the need for more comprehensive inquiries into mechanisms beyond just gender and modes of collaboration. Aspects like societal biases, variations in funding, career-driven pressures, and ethical dilemmas warrant a thorough examination. Grasping these intricacies is crucial to comprehend the significant shifts occurring within the scientific community.

# 7. References

Barbour, Boris, et Brandon Stell. 2020. « PubPeer: Scientific Assessment without Metrics ». in *Gaming the Metrics: Misconduct and Manipulation in Academic Research*. MIT Press.

Blatt, Michael R. 2015. « Vigilante Science ». *Plant Physiology* 169(2):907-9. doi: 10.1104/pp.15.01443.

Bordignon, Frederique. 2020. « Self-Correction of Science: A Comparative Study of Negative Citations and Post-Publication Peer Review ». *Scientometrics* 124(2):1225-39. doi: 10.1007/s11192-020-03536-z.

Bouter, Lex M., Joeri Tijdink, Nils Axelsen, Brian C. Martinson, et Gerben ter Riet. 2016. « Ranking major and minor research misbehaviors: results from a survey among participants of four World Conferences on Research Integrity ». *Research Integrity and Peer Review* 1(1):17. doi: 10.1186/s41073-016-0024-5.

Cokol, Murat, Fatih Ozbay, et Raul Rodriguez-Esteban. 2008. « Retraction rates are on the rise ». *EMBO reports* 9(1):2-2. doi: 10.1038/sj.embor.7401143.

Cole, Jonathan R., et Harriet Zuckerman. 1984. « PERSISTENCE AND CHANGE IN PATTERNS OF PUBLICATION OF MEN AND WOMEN SCIENTISTS ».

Decullier, Evelyne, et Hervé Maisonneuve. 2021. « Retraction according to gender: A descriptive study ». *Accountability in Research* 0(0):1-6. doi: 10.1080/08989621.2021.1988576.

Decullier, Evelyne, et Florence Sèdes. 2022. « Méconduites académiques : exploration d'une distinction potentielle de genres ». in *Actes du 2ème Colloque de l'IRAFPA – Article 18 – Responsable Academia*. Portugal.

Dubois, Michel, et Catherine Guaspare. 2019. « "Is someone out to get me?" : la biologie moléculaire à l'épreuve du Post-Publication Peer Review ». *Zilsel* 6(2):164-92. doi: 10.3917/zil.006.0164.

Fang, Ferric C., Joan W. Bennett, et Arturo Casadevall. 2013. « Males Are Overrepresented among Life Science Researchers Committing Scientific Misconduct ». *mBio* 4(1):e00640-12. doi: 10.1128/mBio.00640-12.

Fang, Ferric C., et Arturo Casadevall. 2011. « Retracted Science and the Retraction Index ». *Infection and Immunity* 79(10):3855-59. doi: 10.1128/iai.05661-11.

Ferguson, Cat, Adam Marcus, et Ivan Oransky. 2014. « Publishing: The Peer-Review Scam ». *Nature* 515(7528):480-82. doi: 10.1038/515480a.

Kaatz, Anna, Paul N. Vogelman, et Molly Carnes. 2013. « Are Men More Likely than Women To Commit Scientific Misconduct? Maybe, Maybe Not ». *mBio* 4(2):10.1128/mbio.00156-13. doi: 10.1128/mbio.00156-13.

Larivière, Vincent, Chaoqun Ni, Yves Gingras, Blaise Cronin, et Cassidy R. Sugimoto. 2013. « Bibliometrics: Global Gender Disparities in Science ». *Nature* 504(7479):211-13. doi: 10.1038/504211a.

Lincoln, Anne E., Stephanie Pincus, Janet Bandows Koster, et Phoebe S. Leboy. 2012. « The Matilda Effect in Science: Awards and Prizes in the US, 1990s and 2000s ». *Social Studies of Science* 42(2):307-20. doi: 10.1177/0306312711435830.

Maddi, Abdelghani, et Yves Gingras. 2021. « Gender Diversity in Research Teams and Citation Impact in Economics and Management ». *Journal of Economic Surveys* 0(0):1-24. doi: https://doi.org/10.1111/joes.12420.

Nicholas, David, Eti Herman, Abdullah Abrizah, Blanca Rodríguez-Bravo, Cherifa Boukacem-Zeghmouri, Anthony Watkinson, Marzena Świgoń, Jie Xu, Hamid R. Jamali, et Carol Tenopir. 2023. « Never Mind Predatory Publishers… What about 'Grey' Publishers? » *Profesional de La Información* 32(5). doi: 10.3145/epi.2023.sep.09.

Ortega, José Luis. 2021. « The Relationship and Incidence of Three Editorial Notices in PubPeer: Errata, Expressions of Concern, and Retractions ». *Learned Publishing* 34(2):164-74. doi: 10.1002/leap.1339.

Ortega, José Luis. 2022. « Classification and Analysis of PubPeer Comments: How a Web Journal Club Is Used ». *Journal of the Association for Information Science and Technology* 73(5):655-70. doi: 10.1002/asi.24568.

Ortega, José-Luis et Lorena Delgado-Quirós. 2023. « How Do Journals Deal with Problematic Articles. Editorial Response of Journals to Articles Commented in PubPeer ». *Profesional de La Información* 32(1). doi: 10.3145/epi.2023.ene.18.

Pfeifer, Mark P., et Gwendolyn L. Snodgrass. 1990. « The Continued Use of Retracted, Invalid Scientific Literature ». *JAMA* 263(10):1420-23. doi: 10.1001/jama.1990.03440100140020.

Pinho-Gomes, Ana-Catarina, Carinna Hockham, et Mark Woodward. 2023. « Women's Representation as Authors of Retracted Papers in the Biomedical Sciences ». *PLOS ONE* 18(5):e0284403. doi: 10.1371/journal.pone.0284403.

Reskin, Barbara F. 1978. « Scientific Productivity, Sex, and Location in the Institution of Science ». *American Journal of Sociology* 83(5):1235-43. doi: 10.1086/226681.

Ribeiro, Mariana D., Jesus Mena-Chalco, Karina de Albuquerque Rocha, Marlise Pedrotti, Patrick Menezes, et Sonia M. R. Vasconcelos. 2023. « Are female scientists underrepresented in self-retractions for honest error? » *Frontiers in Research Metrics and Analytics* 8:1064230. doi: 10.3389/frma.2023.1064230.

Rossiter, Margaret W. 1993. « The Matthew Matilda Effect in Science ». *Social Studies of Science* 23(2):325-41. doi: 10.1177/030631293023002004.

Rossiter, Margaret W. 1997. « Which Science? Which Women? » *Osiris* 12(1):169-85. doi: 10.1086/649272.

Rossiter, Margaret W. 2012. *Women Scientists in America: Forging a New World Since 1972*. JHU Press.

Satalkar, Priya, et David Shaw. 2019. « How do researchers acquire and develop notions of research integrity? A qualitative study among biomedical researchers in Switzerland ». *BMC Medical Ethics* 20(1):72. doi: 10.1186/s12910-019-0410-x.

Steen, R. Grant. 2011. « Retractions in the Scientific Literature: Is the Incidence of Research Fraud Increasing? » *Journal of Medical Ethics* 37(4):249-53. doi: 10.1136/jme.2010.040923.

Steen, R. Grant, Arturo Casadevall, et Ferric C. Fang. 2016. *Why has the number of scientific retractions increased?* Washington, DC, US: American Psychological Association.

Sugimoto, Cassidy R., et Vincent Larivière. 2023. *Equity for Women in Science: Dismantling Systemic Barriers to Advancement*. Harvard University Press.

Teixeira da Silva, Jaime A. 2018. « The opacity of the PubPeer Foundation: what PubPeer's "About" page tells us ». *Online Information Review* 42(2):282-87. doi: 10.1108/OIR-06-2017-0191.

Tower, Greg, Julie Plummer, et Brenda Ridgewell. 2007. « A Multidisciplinary Study Of Gender-Based Research Productivity In The Worlds Best Journals ». *Journal of Diversity Management (JDM)* 2(4):23-32. doi: 10.19030/jdm.v2i4.5020.

UNESCO. 2015. *UNESCO Science Report: Towards 2030*. édité par F. Schlegel. Paris: UNESCO Publ.

Wadman, Meredith. 2005. « One in Three Scientists Confesses to Having Sinned ». *Nature* 435(7043):718-19. doi: 10.1038/435718b.

Wais, Kamil. 2016. « Gender Prediction Methods Based on First Names with genderizeR. » *The R Journal* 8.

West, Jevin D., Jennifer Jacquet, Molly M. King, Shelley J. Correll, et Carl T. Bergstrom. 2013. « The Role of Gender in Scholarly Authorship ». *PLOS ONE* 8(7):e66212. doi: 10.1371/journal.pone.0066212.

Wray, K. Brad, et Line Edslev Andersen. 2018. « Retractions in Science ». *Scientometrics* 117(3):2009-19. doi: 10.1007/s11192-018-2922-4.

Zuckerman, Harriet, Jonathan R. Cole, et John T. Bruer. 1991. *The outer circle: Women in the scientific community*. New York, NY, US: W. W. Norton & Company.